\documentclass[a4paper, 10pt, conference]{ieeeconf}      % Use this line for a4
\usepackage{FG2024}
\usepackage{amsmath} 
\FGfinalcopy % *** Uncomment this line for the final submission

\IEEEoverridecommandlockouts                              % This command is only
\usepackage{graphicx}
\usepackage{array}
\usepackage{multirow}

\usepackage{algorithm}
\usepackage{algpseudocode}
\usepackage{subcaption}
\usepackage[hidelinks]{hyperref}
\usepackage{tikz}
\usepackage{pgfplots}
\def\FGPaperID{312} % *** Enter the FG2024 Paper ID here

\title{\LARGE \bf
Enhancing Privacy in Face Analytics \\ Using Fully Homomorphic Encryption
}

\author{\parbox{16cm}{\centering
    {\large $^1$ $^2$}\\
    {\normalsize
    $^1$ \\
    $^2$ }}
    \thanks{}% <-this % stops a space
}

\author{\parbox{16cm}{\centering
    {\large Bharat Yalavarthi$^{1,*}$ Arjun Ramesh Kaushik$^{1,*}$ Arun Ross$^2$ Vishnu Boddeti$^2$ Nalini Ratha$^1$}\\
    {\normalsize
    $^1$ University at Buffalo, New York, USA\\
    $^2$ Michigan State University, Michigan, USA\\
   \{byalavar, kaushik3, nratha\}@buffalo.edu, \{rossarun, vishnu\}@msu.edu
    \thanks{* First two authors contributed equally to this work.}}}
}

\usepackage{fancyhdr}
\begin{document}

\ifFGfinal
\thispagestyle{empty}
\pagestyle{empty}
\else
%\author{Anonymous FG2024 submission\\ Paper ID \FGPaperID \\}
\pagestyle{plain}
\fi
\maketitle
\thispagestyle{fancy}

%%%%%%%%%%%%%%%%%%%%%%%%%%%%%%%%%%%%%%%%%%%%%%%%%%%%%%%%%%%%%%%%%%%%%%%%%%%%%%%%
\begin{abstract}

Modern face recognition systems utilize deep neural networks to extract salient features from a face. These features denote embeddings in latent space and are often stored as templates in a face recognition system. These embeddings are susceptible to data leakage and, in some cases, can even be used to reconstruct the original face image. To prevent compromising identities, template protection schemes are commonly employed. However, these schemes may still not prevent the leakage of soft biometric information such as age, gender and race. To alleviate this issue, we propose a novel technique that combines Fully Homomorphic Encryption (FHE) with an existing template protection scheme known as PolyProtect. We show that the embeddings can be compressed and encrypted using FHE and transformed into a secure PolyProtect template using polynomial transformation, for additional protection. We demonstrate the efficacy of the proposed approach through extensive experiments on multiple datasets. Our proposed approach ensures irreversibility and unlinkability, effectively preventing the leakage of soft biometric attributes from face embeddings without compromising recognition accuracy. 

\end{abstract}

%%%%%%%%%%%%%%%%%%%%%%%%%%%%%%%%%%%%%%%%%%%%%%%%%%%%%%%%%%%%%%%%%%%%%%%%%%%%%%%%
\section{Introduction}

\begin{figure*}[htbp]
\centerline{\includegraphics[width=0.85\textwidth]{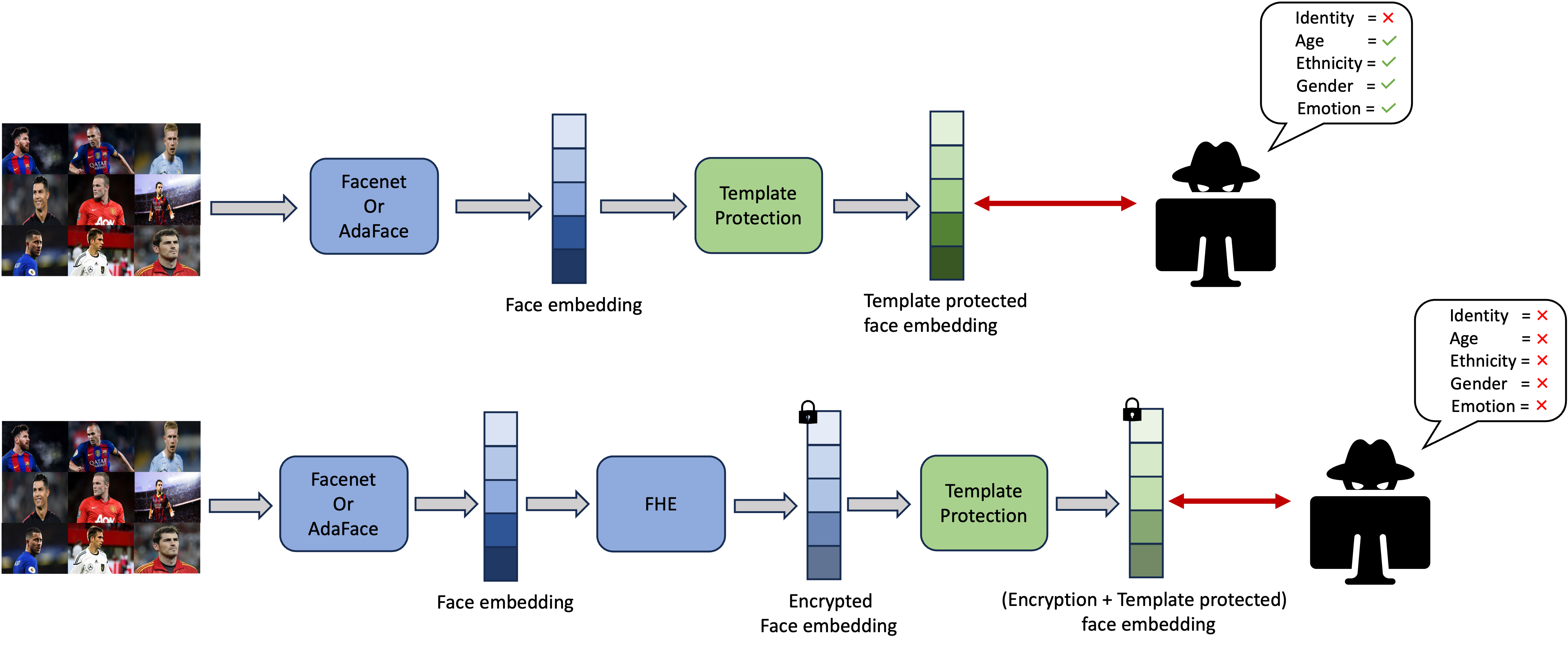}}
\caption{An overview of our contribution in protecting face embeddings from leaking soft biometric attributes - Age, Gender, Ethnicity (FHE - Fully Homomorphic Encryption).}
\label{intro}
\end{figure*}

Face recognition entails the extraction of features from face images and comparing them to either validate a claimed identity (``verification") or determine an identity (``identification") \cite{jain2011introduction}. Recent advancements in deep neural networks and AI have resulted in the development of powerful face recognition systems \cite{taigman2014deepface, Kim_2022_CVPR,handbook-face_2014} that can be deployed in a wide range of applications such as personalized services, law enforcement, border security, and smartphone access \cite{jain50years2016}. However, this development has also raised questions about privacy accorded to subjects and the security of the templates (such as embeddings) stored in a face recognition application \cite{BiometricsReveal2016}. Even as ethical concerns attendant to the technology are being rightfully discussed in public forums, it is necessary for the technology itself, on the one hand, and its users, on the other hand, to embrace measures that can enhance privacy and security while mitigating potential biases \cite{almeida2022ethics}. Otherwise, the technology runs the risk of being overwhelmed by restrictive legislation \cite{chinaGovtFR, FABan} that can stifle the benefits of this technology in solving egregious crimes \cite{westlake2022developing}.

To address some of the privacy challenges associated with {\em face embeddings} stored as {\em templates}, we propose an approach in this paper that employs a polynomial transformation on homomorphically encrypted face embeddings. Using Fully Homomorphic Encryption (FHE) in our proposed method ensures that the face recognition result is only disclosed to authorized parties with the secret key for homomorphic encryption, and the face embeddings themselves are secured through encryption during the recognition process. We show through our experiments that using FHE also prevents leakage of soft biometrics (e.g., age, gender/sex, race/ethnicity) from face embeddings.
%\footnote{
(It is necessary to point out that there is a difference between {\em race} and {\em ethnicity}, as well as {\em sex} and {\em gender}. In this paper, however, we use these terms interchangeably).
%} 

PolyProtect \cite{base}, a template protection scheme, transforms face embeddings into more secure templates using multivariate polynomials with user-specific parameters. Our research demonstrates a symbiotic relationship between FHE and PolyProtect in ensuring optimal security measures. The synergy between these two techniques is essential, as each contributes unique strengths that, when combined, establish a robust security framework. According to the inversion attack analysis conducted by PolyProtect \cite{base}, the risk of reversibility is notably high (30 - 99\%) when an attacker possesses multiple (more than 4) templates of the same face. Even with the compromise of a single template, there is a 95\% likelihood of reversibility when an $overlap$ of 4 or greater is employed. However, PolyProtect introduces a tradeoff between accuracy and security, wherein increasing the $overlap$ parameter enhances accuracy in face recognition and analysis tasks but concurrently diminishes template security. To address this challenge, we implement a strategy of encrypting the face embeddings using FHE and then applying the PolyProtect template with high overlap. This proactive step ensures that even under the full disclosure threat model, irreversibility, unlinkability and prevention of soft biometrics leakage are ensured without compromising identification performance.

Conversely, the integration of PolyProtect into our approach serves as a countermeasure against threats targeting compromised FHE systems, such as secret key leaks and passive attacks outlined in \cite{FHEThreat}. The identified risks pertain to FHE systems engaged in machine learning computations, such as mean and variance calculations. By combining both FHE and PolyProtect techniques (Fig. \ref{intro}), our approach offers a more comprehensive and resilient privacy solution compared to relying on either method in isolation.

\section{Threat Model}
We use the term {\em facial analytics} to refer to the process of deducing semantic information from a face image. This could include sensitive information such as age, gender, ethnicity, and health \cite{HealthCues2022Ross} -- sometimes known as soft biometrics. The possibility of deducing soft biometric cues from face images or their embeddings using automated techniques is a source of concern. These automated techniques can be machine learning models such as SVMs or deep neural networks (DNNs). For example, a face image or its embedding can be ``stolen" by hackers and various soft biometrics can be derived from them thereby revealing sensitive information.

In our work, we presume that the face embedding is provided in an encrypted form. Our goal is to ensure that the encrypted embedding does not reveal any soft biometric information to unauthorized users. Note that the threat remains unchanged even if the parameters of the models used for extracting soft biometric information (e.g., weights of a DNN) are encrypted. 

%While we do not assume that the weights of analytics engines are not encrypted, it wouldn't change much in terms of the threat if they could be encrypted. 
We consider the most challenging threat model according to ISO/IEC 30316, which is the full disclosure model, where the attacker possesses complete knowledge of the PolyProtect method \cite{base}, including its algorithm, number of embedding elements ($m$), user-specific parameters (e.g., $C$, $overlap$, and $E$), and one or more PolyProtected templates corresponding to a face embedding. In addition, we assume that the public key used in FHE is available but not the private key. If the embeddings are {\em not} encrypted, the hacker can infer soft biometric information from the PolyProtect template as we show in Table \ref{Accuracy}. 
%Does this sentence convey revocability??? 
%\color{black} On the other hand, our proposed approach does not protect soft-biometric leakage in imbalanced datasets.

\section{FHE Basics}
Encryption is the process where plain-text data is encrypted into ciphertext using a secret key and a cryptographic algorithm. Only authorized entities with a private key can decrypt the ciphertext back to the plaintext. Encryption is essential for protecting sensitive data from unauthorized access or modification. 
Homomorphic Encryption (HE) is a cryptographic system that permits certain computations to be performed on encrypted data without requiring decryption \cite{rivest1978}. In this system, we have public ($pk$) and secret ($sk$) keys, encryption ($E$) and decryption ($D$) mechanisms, and plaintext values $x$ and $y$. When $x$ and $y$ are encrypted as $x' = E(x, pk)$ and $y' = E(y, pk)$, respectively, a cryptosystem is considered homomorphic with respect to a chosen operator (e.g., addition or multiplication), denoted as $\circ$, if we can find another operator $\bullet$ such that $x \circ y = D(x' \bullet y', sk)$. This means that we can conduct operations on encrypted data and obtain the same result when decrypting using the private secret key.

Specifically, given $c_i = E(x_i, pk), i = 1, 2, \cdots, K$, an FHE scheme allows the computation of $c = g(c_1, c_2, \cdots, c_K)$ such that $D(c, sk) = f(x_1, x_2, \cdots, x_K)$ for any arbitrary function $f$. 

It is essential to note that there are three types of homomorphic encryption schemes \cite{he_taxonomy}: (a) Partial Homomorphic Encryption (PHE) permits addition or multiplication operations. (b) Somewhat Homomorphic Encryption (SHE) allows limited computations on ciphertexts. (c) Fully Homomorphic Encryption (FHE) enables computations on ciphertexts of any depth and complexity.
% \item Leveled Homomorphic Encryption (LHE) supports computations on ciphertexts with limited depth, with the option to increase depth by using multiple levels of encryption.

% \color{red}
% FHE is often achieved by employing a \emph{somewhat homomorphic} (SWHE) or \emph{leveled} HE scheme in combination with a \emph{bootstrapping} or \emph{recryption} technique. 
% The SWHE scheme is capable of supporting computations only up to a preset level of complexity determined by its parameters. This is because the ciphertexts are ``noisy'', the noise keeps growing with each HE computation, and once the noise grows beyond some (parameter-dependent) threshold the ciphertext can no longer be decrypted. This problem is solved using Gentry's bootstrapping technique \cite{gentry2012fully}, which ``refreshes'' the ciphertext and reduces its noise level (at the cost of relying on circular security). However, bootstrapping is a computationally expensive and time-consuming operation. Therefore, for practical feasibility, the number of bootstrapping operations should be kept at a minimum and possibly avoided.
% \color{black}

Numerous FHE systems have been introduced, including the BFV, BGV, and CKKS schemes \cite{FHEACM}. The BFV and BGV schemes enable vector operations involving integers, while the CKKS scheme facilitates floating-point operations. These schemes achieve Single Instruction Multiple Data (SIMD) operations by bundling plaintext values into an array and then encrypting them to get ciphertext. In this work we use the HEAAN \cite{HEAAN} library based on the CKKS scheme for FHE computations.

%%%%%%%%%%%%%%%%%%%%%%%%%%%%%%%%%%%%%%%%%%%%%%%%%%%%%%%%%%%%%%%%%%%%%%%%%%%%%%%%
\section{Related Work}
A notable body of literature explores privacy enhancements to soft biometrics at both the image and embedding (template) levels in face recognition. PFRNet \cite{PFRNet} uses an Autoencoder framework to disentangle identity from attribute information to suppress gender information in face embeddings. Similarly, SensitiveNets \cite{morales2020sensitivenets} uses a privacy-preserving neural network that suppresses soft biometrics attributes. The approach adopts an adversarial regularizer, which incorporates a sensitive information removal function into the learning objective. The Multi Incremental Variable Elimination (Multi-IVE) method \cite{mIVE} works by eliminating those feature variables in embeddings that predict soft biometric attributes. Increasing the number of eliminations was shown to decrease the soft biometrics leakage but significantly affected the identification performance. In \cite{mutual}, the authors introduce an adversarial attack approach designed to protect gender information in facial images. The method involves perturbing the image to minimize the estimated mutual information between the feature distribution acquired from a face recognition network and the gender variable. This method reduces gender leakage (prediction accuracy) by an average of 91\% to 80\% across three datasets. Reversible Attribute Privacy Preservation (RAPP) \cite{RAPP} uses a stream cipher to determine the sensitive attributes that have to be concealed with a user-defined password; it supports recovering the original attributes. It also uses an attribute adversarial network to generate perturbed images that conceal various attributes while retaining the utility of face verification. However, the identification performance is negatively impacted. The authors work around this challenge by reducing the number of features being concealed and the intensity of concealment. PrivacyNet \cite{PrivacyNet} is another technique to impart soft biometric privacy to face images while preserving recognition capabilities via image perturbation using a GAN-based Semi-Adversarial Network (SAN). PrivacyNet also allows a person to choose the specific facial attributes to be obfuscated while allowing the other attributes to be extracted. One of the drawbacks of this image perturbation technique is that it sometimes does not generate realistic images and cannot conceal soft biometric features from a human observer.

Although current approaches mitigate the leakage of soft biometric attributes, they do not suppress it to the level of a ``random guess". In our work, we show that through the use of FHE, we can restrict this leakage in face embeddings to a level that is equal to or lower than that of a random guess. In \cite{FHEThreat}, the authors show the susceptibility of privacy enhancement techniques such as homomorphic encryption. We address this susceptibility by employing a template protection scheme in addition to homomorphic encryption. Further, FHE encryption offers a stricter theoretical guarantee than existing methods for the security of soft biometrics.

FHE has been used in prior work for securing face recognition. Boddeti \cite{boddeti2018secure} proposed encrypting the face embeddings and performing face matching in the FHE domain. Batching and dimensionality reduction techniques are also explored to balance face-matching accuracy and computational complexity. In \cite{engelsma2022hers}, the authors introduce an efficient approach for searching encrypted probe images against a large gallery, using fixed-length representations. In \cite{ppFHE}, the authors proposed a time-efficient and space-efficient face matching in the FHE domain for securing face templates. Our approach stands out from previous work by integrating a template protection scheme, a compression technique, and FHE to enhance the security of face templates. Moreover, we conducted thorough experiments to assess their efficacy in mitigating the leakage of soft biometrics.  

%Unlike ours which restricts the soft biometrics leakage to the level of random guessing, this framework does not fully suppress the attribute information and suffers from loss in identification performance.     

%Although there are some of these approaches that suppress facial attributes to protect privacy, Our work achieves the privacy of the soft biometrics in face representations by FHE encryption.  FHE encryption is a stricter form of security for these features than suppressing them.

%\textbf{Polyprotect} \cite{base} has been developed as a contribution towards preserving the privacy of face recognition systems by mapping sensitive face embeddings to more secure representations through homomorphic encryption. PolyProtectuses multivariate polynomials with user-defined parameters to derive the representations. The paper evaluates their work in a cooperative-user mobile verification scenario on two open-source face recognition systems. It goes on to prove the irreversibility and unlinkability of the templates, as well.

\section{Template Protection}

Amongst the many existing protection templates, we have adopted \textbf{PolyProtect} \cite{base} to showcase the benefits of our work. Let $V = [v_1, v_2, ..., v_n]$ denote an $n$-dimensional real-number face embedding. PolyProtect maps $V$ to another real-numer feature vector, $P = [p_1, p_2, ..., p_k]$ (where $k<n$). $P$ is the PolyProtected template of $V$. $m$ (where $m<<n$) consecutive elements from $V$ are mapped to single elements in $P$ via a polynomial equation of coefficients, $C = [c_1, c_2, ..., c_m]$ and exponents, $E = [e_1, e_2, ..., e_m]$. $C$ and $E$ are user-defined, non-zero, and distinct for each user of the face recognition system. The first $m$ elements of $V$ are mapped to $p_1$ as :
\begin{align}
    p_1 &= c_1v_{1}^{e_1} + c_2v_{2}^{e_2} + ... + c_mv_{m}^{e_m} 
%\vspace{-0.2in}
\end{align}

Another important user-defined parameter is $overlap$, which defines the number of common elements from $V$ used in successive values in $P$. When $overlap = 0$, the elements of $V$ in each set are unique. The minimum and maximum values for $overlap$ are $0$ and $m-1$, respectively. The mapping for $p_2$ for overlaps $0$ and $m-1$, respectively, are as follows:
\begin{align}
     p_2 &= c_1v_{m+1}^{e_1} + c_2v_{m+2}^{e_2} + ... + c_mv_{m+m}^{e_m}\\
     p_2 &= c_1v_{2}^{e_1} + c_2v_{3}^{e_2} + ... + c_mv_{m+1}^{e_m}
\end{align}

The authors of PolyProtect \cite{base} have also performed an extensive survey on a number of existing Biometric Template Protection (BTP) methodologies and evaluated them based on recognition accuracy, irreversibility, and unlinkability \cite{BTPSurvey}. According to the survey, none of the existing BTP methodologies before PolyProtect \cite{base} satisfy all three criteria - recognition accuracy, irreversibility, and unlinkability.

\section{Ablation Study}

\begin{figure*}
    % \centering
    % \hspace{-0.5cm}
    \includegraphics[width=0.66\textwidth]{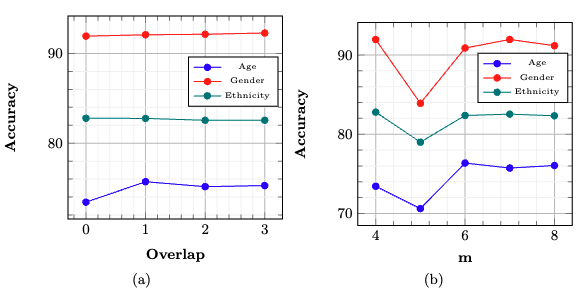}
    \includegraphics[width=0.33\textwidth]{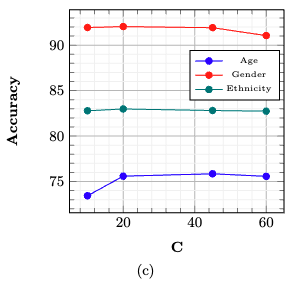}
    \caption{Ablation study with the PolyProtect parameters shows soft biometrics leakage in different settings. (a) Overlap.  (b) m - length of polynomial coefficients/exponents. (c) [-C, C] - range of polynomial coefficients.}
    \label{Fig:ablation}
\end{figure*}

\subsection{Variation of user-defined parameters ($C$, $m$, $overlap$)}
We conducted experiments with various user parameters in PolyProtect to investigate their impact on the leakage of soft biometrics. As $overlap$ increased from 0 to 3, there was a notable increase in age leakage, particularly for values $overlap>0$. However, gender and ethnicity showed consistent levels of leakage across different values of $overlap$ (Fig. \ref{Fig:ablation}(a)).  As the overlap increases, the amount of embedding information retained after the PolyProtect transformation also increases, potentially resulting in a higher risk of soft biometric leakage. 
In PolyProtect, the parameter $m$ dictates the number of terms in the polynomial. At $m=6$, we observed maximum leakage for age, whereas for gender and ethnicity, maximum leakage occurred at $m=7$. Conversely, selecting $m=5$ minimized leakage across all three soft biometric attributes (Fig. \ref{Fig:ablation}(b)). Additionally, the parameter $C$, which defines the range of values for the polynomial coefficients [-$C$, $C$], was varied from 10 to 60 in our study. The leakage of ethnicity remained relatively stable across all tested values of $C$, while age exhibited an increase from $C$ = 10 to 20, and gender showed a slight decrease from $C$ = 45 to 60 (Fig. \ref{Fig:ablation}(c)).

%\vspace{-0.18cm}

\subsection{Summation of Ciphertext elements}
Implementing PolyProtect, cosine similarity, and fully connected layers in FHE requires summing up the elements within a ciphertext. This is not straightforward as we cannot access individual elements of a ciphertext. The three approaches to efficiently achieve the summation are described below:

\subsubsection{\textbf{Naive Rotation}}
This is a brute-force method for summation within a ciphertext where expensive ciphertext rotations are performed $N-1$ times and the running sum is computed until all the elements are covered (Algorithm \ref{alg:a1}).  
% \begin{algorithm}
% \caption{Add Ciphertext Elements Through Left Rotation n-1 times}\label{alg:nRot}
% \begin{algorithmic}
% \Function{NAIVE\_ADD}{Ciphertext $c$, long $N$}
%     \State Ciphertext $c1 \gets c$
%     \While{$N > 0$}
%         \State $LeftRot(c1, 1)$ \Comment{Left Rotates Ciphertext by 1}
%         \State $c \gets Add(c, c1)$ \Comment{Adds two Ciphertexts}
%         \State $N \gets N-1$
%     \EndWhile
%     \State \Return $c$
% \EndFunction
% \end{algorithmic}
% \label{alg:a1}
% \end{algorithm}

\subsubsection{\textbf{Discrete Fourier Transform}}
When the Discrete Fourier Transform (DFT) of a signal is computed, the first value of DFT or the DC component will give the sum of the input signal values. We use this property to calculate the DFT of the ciphertext and get the sum of its values.     

\subsubsection{\textbf{Fold and Add}}
\begin{figure} [!ht]
  \centering
  \begin{tikzpicture}
    \begin{axis}[
      width=0.45\textwidth,
      height=0.35\textwidth,
      xlabel={Ciphertext Size ($2^x$)},
      ylabel={Time Taken (ms)},
      font=\bfseries,
      legend pos=north west,
      legend style ={font=\small},
      xmin=2, xmax=2048,
      xmode=log,
      xtick={2, 4, 8, 16, 32, 64, 128, 256, 512, 1024, 2048},
      xticklabels={1, 2, 3, 4, 5, 6, 7, 8, 9, 10, 11},
      grid=both,
      grid style={line width=.1pt, draw=gray!10},
      major grid style={line width=.2pt,draw=gray!50},
      minor tick num=4,
    ]

    % Scatter plots and lines
    \addplot[mark=*, blue, line width = 0.5pt] table[x=Size, y=Time] {
      Size Time
      2 21.17
      4 29.55
      8 44.66
      16 77.24
      32 140.79
      64 307.74
      128 634.44
      256 1310.53
      512 2957.69
      1024 4732.74
      2048 9088.11
    };
    \addlegendentry{Naive Rotation}

    \addplot[mark=*, red, line width = 0.5pt] table[x=Size, y=Time] {
      Size Time
      2 9.00
      4 34.51
      8 41.37
      16 58.51
      32 80.84
      64 104.00
      128 130.81
      256 137
      512 256.06
      1024 197.82
      2048 163.92
    };
    \addlegendentry{DFT}

    \addplot[mark=*, teal, line width = 0.5pt] table[x=Size, y=Time] {
      Size Time
      2 13.51
      4 26.09
      8 34.99
      16 46.63
      32 52.26
      64 72.91
      128 88.17
      256 95.93
      512 128.14
      1024 121.47
      2048 155.35
    };
    \addlegendentry{Fold and Add}

    \end{axis}
  \end{tikzpicture}
  \caption{Execution time to compute summation of elements within a ciphertext.}
  \label{sum}
\end{figure}
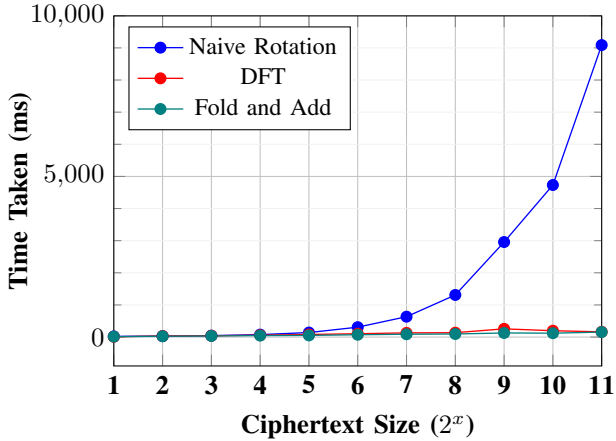

This is a more efficient version of the naive rotation method described earlier, which can be visualized as iteratively folding the array into half and adding the corresponding folded parts $\log_{2} N - 1$ times as described in algorithm \ref{alg:a2}. Fig. \ref{sum} shows that the $Fold$ $and$ $Add$ method was the fastest among the three with a significant speedup than the naive rotation method, especially for large ciphertext sizes.

\begin{algorithm}
\caption{Add Ciphertext Elements Through Left Rotation n-1 times}\label{alg:nRot}
\begin{algorithmic}
\Function{NAIVE\_ADD}{Ciphertext $c$, long $N$}
    \State Ciphertext $c1 \gets c$
    \While{$N > 0$}
        \State $LeftRot(c1, 1)$ \Comment{Left Rotates Ciphertext by 1}
        \State $c \gets Add(c, c1)$ \Comment{Adds two Ciphertexts}
        \State $N \gets N-1$
    \EndWhile
    \State \Return $c$
\EndFunction
\end{algorithmic}
\label{alg:a1}
\end{algorithm}
\vspace{-0.1cm}
\begin{algorithm}
\caption{Add Ciphertext Elements Through Left Rotation $\log_{2} N$ times}\label{alg:n/2Rot}
\begin{algorithmic}
\Function{FOLD\_ADD}{Ciphertext $c$, long $N$}
    \State $k \gets \lceil \log_{2} N \rceil$ 
    \State $i \gets k - 1$ 
    \While{$i > 0$}
        \State Ciphertext $c1 \gets LeftRot(c, 2^{i})$ 
        \State $c \gets Add(c, c1)$ 
        \State $i \gets i-1$
    \EndWhile
    \State \Return $c$
\EndFunction
\end{algorithmic}
\label{alg:a2}
\end{algorithm}

% \begin{figure}[!ht]
%       \centering
%       \includegraphics[width=0.5\textwidth]{figures/summation.png}
%       \caption{Time taken to compute summation of elements within a ciphertext.}
%       \label{sum}
%    \end{figure}

\vspace{-0.15cm}
\subsection{Polynomial approximation of inverse square root}

\begin{figure}[!ht]
    % \centering
    \hspace{-0.5cm}
    \includegraphics[width=0.5\textwidth]{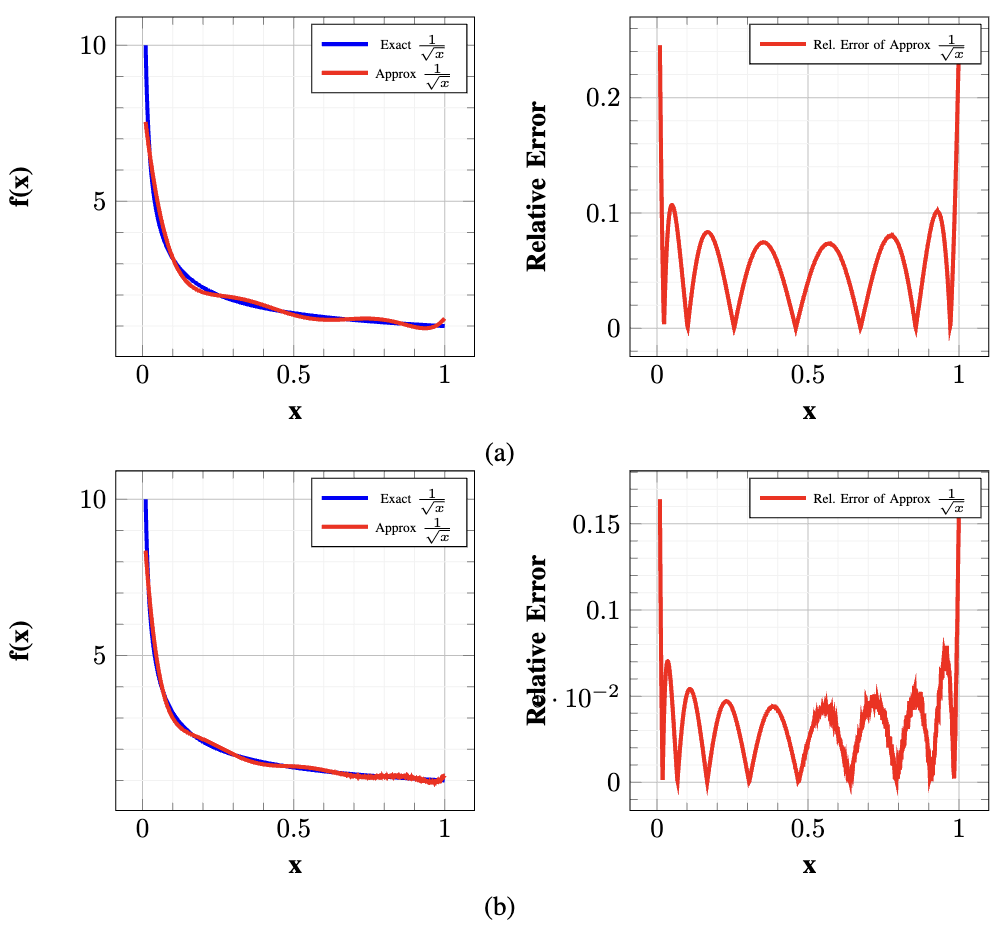}
    \caption{Inverse square root function. (a) 6-degree polynomial approximation and (b) 8-degree polynomial approximation and their relative error over 2000 random points in the range (0,1].}
    \label{fig:inv_sq_root}
\end{figure}

Since computing cosine similarity requires inverse square root function, we use a polynomial approximation of the inverse square root function in the encrypted domain owing to the limitations of FHE in implementing non-linear functions. We restrict the input to the range (0, 1] to achieve a closer approximation. The performance of this approximation is measured through the relative error of 2000 random points in the range (0, 1]. We consider both 6-degree and 8-degree polynomials as a tradeoff between computational depth and accuracy, as in Fig. \ref{fig:inv_sq_root}. 
\begin{table*}[!h]
\caption{Statistics of the CelebSet Dataset (80 Identities).}
\label{CelebsetData}
\begin{center}
\begin{tabular}{|c|c|c|c|c|c|c|c|c|c|c|c|}
\hline
\multicolumn{2}{|c|}{\textbf{Gender}} & \multicolumn{4}{|c|}{\textbf{Age}} & \multicolumn{4}{|c|}{\textbf{Ethnicity}} \\
\hline
\textbf{Males} & \textbf{Females} & \textbf{0-22} & \textbf{23-40} & \textbf{41-59} & \textbf{60+} & \textbf{Hispanic} & \textbf{White} & \textbf{Black} & \textbf{Asian}\\
\hline
38,080 & 34,409 & 5,279 & 43,357 & 22,781 & 1,072 & 738 & 57,873 & 13,414 & 464\\ 
\hline
52.50\% & 47.50\% & 7.28\% & 59.82\% & 31.43\% & 1.47\% & 1.01\% & 79.85\% & 18.50\% & 0.64\%\\ 
\hline
\end{tabular}
\end{center}
\end{table*}

%\vspace{-0.6cm}
\section{Proposed Approach}
The proposed methodology begins with the reception of encrypted face templates from users. These templates are generated through a pre-trained face recognition model and encrypted locally at the user's end using Fully Homomorphic Encryption (FHE), employing a private key exclusively known to the user. Subsequently, a template protection algorithm such as PolyProtect is applied to augment the security of the encrypted template, forming an additional layer of protection. The resulting encrypted PolyProtect template can be securely stored at the user's end and utilized as needed for tasks such as face identification or the extraction of soft biometrics, including gender, age, and ethnicity. The outcomes of these tasks are transmitted to the user in encrypted form and can only be decrypted using the user's private key.
\subsection{Dataset}
We have performed our experiments on CelebSet \cite{celebset} and Balanced Faces in the Wild (BFW) \cite{bfw}. The statistics of these datasets have been detailed in Tables \ref{CelebsetData}, \ref{BFWData}. As the BFW dataset lacks age annotations for its face images, we utilized a pre-trained model trained on the CelebSet dataset to predict the ages of the BFW images. These predicted ages were then employed in our experiments.

\begin{table*}
\caption{Statistics of the BFW Dataset (100 Identities).}
\label{BFWData}
\begin{center}
\begin{tabular}{|c|c|c|c|c|c|c|c|c|c|c|c|}
\hline
\multicolumn{2}{|c|}{\textbf{Gender}} & \multicolumn{6}{|c|}{\textbf{Predicted Age}} & \multicolumn{4}{|c|}{\textbf{Ethnicity}} \\
\hline
\textbf{Males} & \textbf{Females} & \textbf{0-4} & \textbf{5-12} & \textbf{13-19} & \textbf{20-39} & \textbf{40-59} & \textbf{60+} & \textbf{Indian} & \textbf{White} & \textbf{Black} & \textbf{Asian}\\ 
\hline
10,000 & 10,000 & 0 & 50 & 16,326 & 3,612 & 12 & 0 & 5,000 & 5,000 & 5,000 & 5,000\\ 
\hline
50\% & 50\% & 0\% & 0.25\% & 81.63\% & 18.06\% & 0.06\% & 0\% & 25\% & 25\% & 25\% & 25\%\\ 
\hline
\end{tabular}
\end{center}
\end{table*}

\subsection{Embedding Compression}
It is commonly believed that compressing embeddings can improve privacy leakage. We explore embedding compression through \textbf{Matryoshka Representation Learning (MRL}) \cite{matryoshka} as a technique to improve privacy, in addition to template protection and encryption. MRL is an innovative approach that enhances representation learning by encoding information at various granularities within a single embedding. MRL achieves this by capturing details at different levels of abstraction, enabling the model to adapt to downstream tasks efficiently. The method focuses on learning coarse-to-fine representations, ensuring that high-level, generalized features are initially captured and progressively refined with finer details. MRL stands out by delivering representations that are comparable in accuracy and rich in information when compared to independently trained low-dimensional representations, making it a promising solution for efficient and effective representation learning.
% \begin{figure}[!ht]
%       \centering
%       \includegraphics[width=0.5\textwidth]{figures/MRL.png}
%       \caption{Performance of Matryoksha Representation Learning(MRL) in extracting features - Identity, Age, Gender, Ethnicity - from different compression dimensions. The experiment has been performed using AdaFace on the CelebSet dataset.}
%       \label{attack}
% \end{figure}

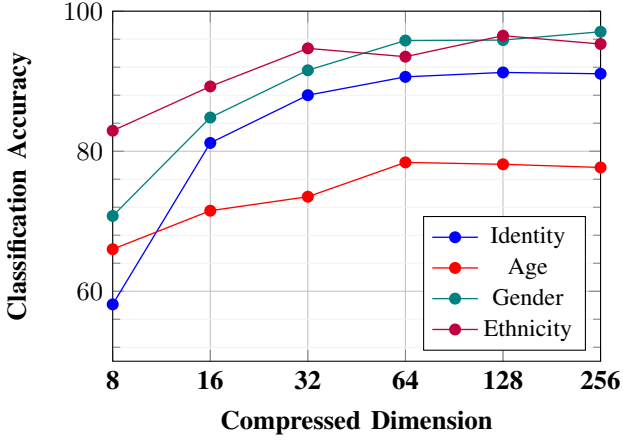
\begin{figure}[!ht]
  \centering
  \begin{tikzpicture}
    \begin{axis}[
      width=0.45\textwidth,
      height=0.35\textwidth,
      xlabel={Compressed Dimension},
      ylabel={Classification Accuracy},
      font=\bfseries,
      legend pos=south east,
      legend style ={font=\small},
      xmin=8, xmax=256,
      ymin=50, ymax=100,
      xmode=log,
      xtick={8, 16, 32, 64, 128, 256},
      xticklabels={8, 16, 32, 64, 128, 256},
      grid=both,
      grid style={line width=.1pt, draw=gray!10},
      major grid style={line width=.2pt,draw=gray!50},
      minor tick num=4,
    ]

    % Scatter plots and lines
    \addplot[mark=*, blue, line width = 0.5pt] table[x=Compression, y=Identity] {
      Compression Identity
      8 58.12
      16 81.18
      32 88
      64 90.62
      128 91.25
      256 91.06
    };
    \addlegendentry{Identity}

    \addplot[mark=*, red, line width = 0.5pt] table[x=Compression, y=Age] {
      Compression Age
      8 66
      16 71.5
      32 73.5
      64 78.4
      128 78.13
      256 77.68
    };
    \addlegendentry{Age}

    \addplot[mark=*, teal, line width=0.5pt] table[x=Compression, y=Gender] {
      Compression Gender
      8 70.75
      16 84.81
      32 91.56
      64 95.81
      128 95.87
      256 97.06
    };
    \addlegendentry{Gender}

    \addplot[mark=*, purple, line width = 0.5pt] table[x=Compression, y=Ethnicity] {
      Compression Ethnicity
      8 82.93
      16 89.25
      32 94.69
      64 93.5
      128 96.5
      256 95.31
    };
    \addlegendentry{Ethnicity}

    \end{axis}
  \end{tikzpicture}
  \caption{Performance of Matryoksha Representation Learning (MRL) in extracting features - Identity, Age, Gender, Ethnicity - from different compressed dimensions. The graph is based on experiments performed using AdaFace on the CelebSet dataset.}
  \label{attack}
\end{figure}

In our research, we leveraged MRL to compress face embeddings obtained from FaceNet/AdaFace, which are originally of 512-dimension, down to 64-dimension. As depicted in Fig. \ref{attack}, we noted a decrease in classification accuracies when the compressed dimensions fell below 64. It is important to note that, since HEAAN supports SIMD operations, compression does not aid in reducing computational depth. 

\subsection{Face Identification}
Given a vector of face embeddings of size $n$, and user-defined parameters $C$, $m$, and $overlap$, we pad the embeddings vector such that it can be split into $m$ small vectors $V_i$ (where $m<<n$). Each $m$-sized vector is encrypted in FHE. We apply the PolyProtect algorithm as described in \cite{base} on each $V_i$, resulting in vector $P_i$. Unlike in the PolyProtect algorithm \cite{base}, we store every PolyProtect mapping as a vector of $m$ elements which are the same, resulting from the sum obtained through $Fold$ $and$ $Add$ (Algorithm \ref{alg:a2}). We have employed the Cosine similarity as a metric to calculate similarity scores in the $1:N$ search for identification. Cosine similarity in FHE is computed using the algorithm described in algorithm \ref{alg:a3}. 

\begin{algorithm}
\caption{Cosine Similarity between ciphertexts $C1$ and $C2$ of size $N$ in FHE}\label{alg:a3}
\begin{algorithmic}
\Function{Cosine\_Distance}{$C1$,$C2$, $N$}
   
    \State $C \gets Mult(C1,C2)$ 
    \State $C \gets FOLD\_ADD(C, N)$
    \State $D1 \gets FOLD\_ADD(Mult(C1,C1), N)$
    \State $D2 \gets FOLD\_ADD(Mult(C2,C2), N)$
    \State $D \gets Mult(D1,D2)$
    \State $C \gets Normalization (C)$ \Comment{Scales all values to the range [0,1]}
    \State $D \gets Normalization (D)$
    \State $D \gets approxInverseSquareRoot(D)$
    \State $C \gets Mult(C,D)$
    
    \State \Return $C$
\EndFunction
\end{algorithmic}
\label{alg:a3}
\end{algorithm}
  
\subsection{Age, Gender and Ethnicity Prediction}

\begin{figure}[!ht]
      \centering
      \includegraphics[width=0.4\textwidth]{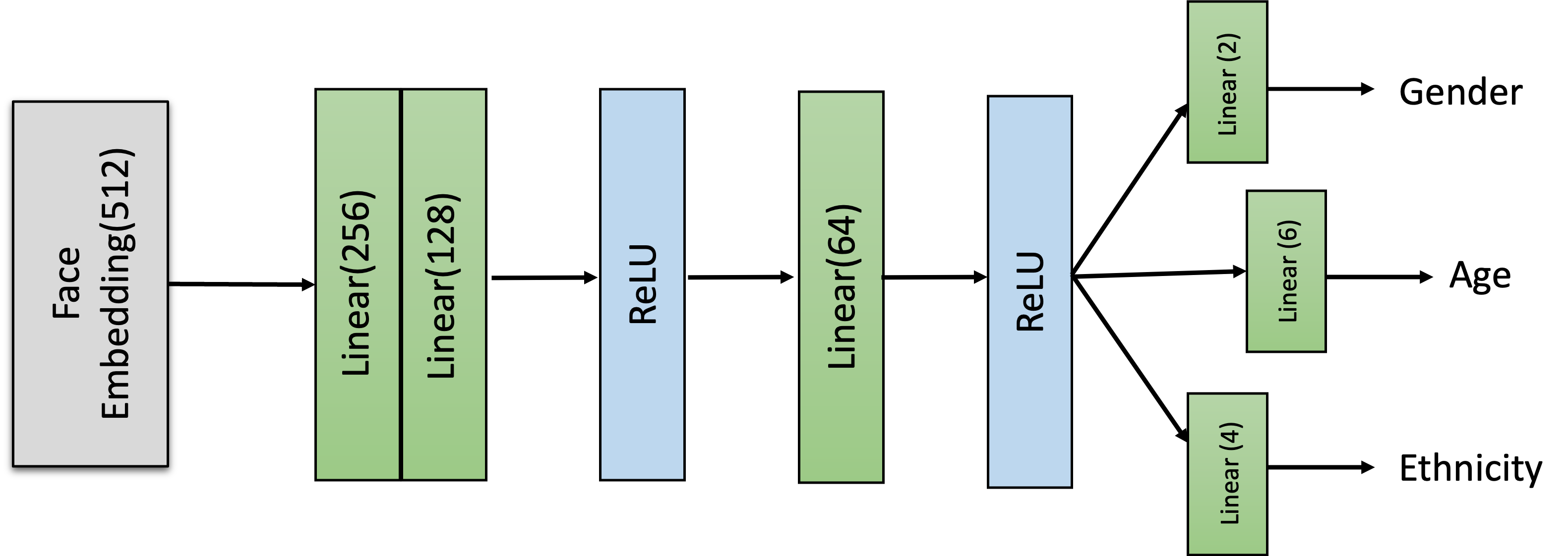}
      \caption{Feed Forward Network used for classifying soft biometrics.}
      \label{net}
   \end{figure}

We perform our experiments using various combinations of embeddings, template protection, compression, and encryption. The embeddings are extracted through FaceNet \cite{facenet} and AdaFace \cite{adaface} models. Our experiments are primarily aimed at determining the identification accuracy and soft biometrics leakage for the described combinations to prove the efficacy of our proposed solution. Our networks (Fig. \ref{net}) have been tuned from the lens of an attacker to achieve maximum leakage of soft biometric features. For cases that require training with encrypted data in the FHE domain, we train an SVM classifier on the ASCII dump of the ciphertexts. 

Apart from a myriad of leakage metrics defined in \cite{leakageMetrics2} and \cite{leakageMetrics3}, we have chosen to define leakage through Suppression Rate (SR) and Privacy Gain (PG) \cite{leakageMetrics}. Privacy Gain is defined as -
\begin{equation}
    \text{PG} = (1 - R_p) - (1-R_o)
\end{equation}
where, $R_o$ and $R_p$ represent the recognition performances on the original data and the privacy-enhanced data, respectively. A positive value of PG signifies enhanced data protection. The ideal value for PG is 1. 

Suppression Rate measures the difference attribute prediction accuracy with and without privacy enhancement, $A_p$ and $A_o$, respectively

\begin{equation}
    \text{SR} = {(A_o - A_p)}/{A_o}
\end{equation}
\vspace{-0.1in}

\begin{table*}
\caption{Face identification and soft biometric classification accuracy (MRL - Matryoksha Representation Learning; FHE - Fully Homomorphic Encryption). Note that the proposed approach retains identification accuracy while successfully reducing soft biometric classification accuracy.}
\label{Accuracy}
\begin{center}
\begin{tabular}{|c|c|c|c|c|c|c|}
\hline
\textbf{Embeddings} & \textbf{Dataset} & \textbf{Template Protection} & \textbf{Identification Accuracy}&\textbf{Gender Accuracy} & \textbf{Age Accuracy} & \textbf{Ethnicity Accuracy}\\
\hline
\multirow{6}{*}{FaceNet} & \multirow{6}{*}{CelebSet} & None & 
99.42\% &98.12\% & 87.68\% & 98.81\%\\ 

& & PolyProtect & 99.42\%&97.35\% & 85.00\% & 98.06\%\\ 

& & MRL & 97.93\% & 98.00\% & 85.87\% & 97.38\%\\ 

& & MRL + PolyProtect &97.00\% & 96.32\% & 87.32\% & 98.44\%\\ 

& & MRL + FHE & 97.83\% & \textbf{52.22\%} & \textbf{6.12\%} & \textbf{8.01\%}\\ 

& & MRL + PolyProtect + FHE & 96.95\% & \textbf{52.22\%} & \textbf{6.12\%} & 8.03\%\\ 
\cline{2-7}

 & \multirow{6}{*}{BFW} & None & 85.06\%&95.25\% & 94.40\% & 91.97\%\\
& & PolyProtect & 85.04\% & 95.07\% & 93.97\% & 91.62\%\\ 

& & MRL &84.42\%& 92.02\% & 93.98\% & 91.65\%\\ 

& & MRL + PolyProtect &84.31\% & 90.00\% & 87.90\% & 87.70\%\\ 

& & MRL + FHE &84.38\%& 49.98\% & 28.00\% & 24.01\%\\ 

& & MRL + PolyProtect + FHE & 84.28\% & \textbf{49.50\%} & \textbf{27.82\%} & \textbf{24.00\%}\\

\hline

\multirow{6}{*}{AdaFace} & \multirow{6}{*}{CelebSet} & None & 99.41\%&96.93\% &90.25\% & 96.31\%\\
& & PolyProtect &99.38\%& 95.75\% & 90.18\% & 96.56\%\\

& & MRL & 97.95\% & 97.63\% & 85.50\% & 98.94\%\\ 

& & MRL + PolyProtect & 97.59\% & 96.82\% & 89.13\% & 98.25\%\\ 

& & MRL + FHE & 97.91\% & \textbf{52.22\%} & 6.12\% & 8.02\%\\ 

& & MRL + PolyProtect + FHE & 97.55\% & \textbf{52.22\%} & \textbf{6.11\%} & \textbf{8.01\%}\\ 
\cline{2-7}

 & \multirow{6}{*}{BFW} & None & 88.55\% & 89.97\% &85.02\% & 81.72\%\\
& & PolyProtect & 88.46\% &84.07\% & 84.62\% & 74.62\%\\

& & MRL & 88.33\% & 87.55\% & 83.83\% & 76.43\%\\ 

& & MRL + PolyProtect & 88.29\% & 85.58\% & 83.66\% & 75.58\%\\ 

& & MRL + FHE & 88.23\% & 49.98\% & 27.97\% & 24.02\%\\ 

& & MRL + PolyProtect + FHE & 88.21\% & \textbf{49.50\%} & \textbf{27.90\%} & \textbf{24.00\%}\\ 

\hline
\end{tabular}
\end{center}
\end{table*}

\begin{table*}
\caption{Privacy Gain across different soft biometric attributes.}
\label{PrivacyGain}
\begin{center}
\begin{tabular}{|c|c|c|c|c|c|c|}
\hline
\textbf{Embeddings} & \textbf{Dataset} & \textbf{Template Protection} & \textbf{Identification Accuracy}&\textbf{Gender Attribute} & \textbf{Age Attribute} & \textbf{Ethnicity Attribute}\\
\hline
\multirow{6}{*}{FaceNet} & \multirow{6}{*}{CelebSet} &  PolyProtect & 99.42\%& 0.72 & 2.68 & 0.75\\ 

& & MRL & 97.93\% & 0.12 & 1.81 & 1.43\\ 

& & MRL + PolyProtect &97.00\% & 1.80 & 0.36 & 0.37\\ 

& & MRL + FHE & 97.83\% & \textbf{45.90} & \textbf{81.56} & \textbf{90.80}\\ 

& & MRL + PolyProtect + FHE & 96.95\% & \textbf{45.90} & \textbf{81.56} & 90.78\\ 
\cline{2-7}

 & \multirow{6}{*}{BFW}  & PolyProtect & 85.04\% & 0.18 & 0.43 & 0.35\\ 

& & MRL &84.42\%& 3.23 & 0.42 & 0.32\\ 

& & MRL + PolyProtect &84.31\% & 5.25 & 6.50 & 4.27\\ 

& & MRL + FHE &84.38\%& 45.27 & 66.40 & 67.96\\ 

& & MRL + PolyProtect + FHE & 84.28\% & \textbf{45.75} & \textbf{66.58} & \textbf{67.97}\\

\hline

\multirow{6}{*}{AdaFace} & \multirow{6}{*}{CelebSet} & PolyProtect &99.38\%& 1.18 & 0.07 & -0.25\\

& & MRL & 97.95\% & -0.7 & 4.75 & -2.63\\ 

& & MRL + PolyProtect & 97.59\% & 0.11 & 1.12 & -1.94\\ 

& & MRL + FHE & 97.91\% & \textbf{44.71} & 84.13 & 88.29\\ 

& & MRL + PolyProtect + FHE & 97.55\% & \textbf{44.71} & \textbf{84.14} & \textbf{88.30}\\ 
\cline{2-7}

 & \multirow{6}{*}{BFW} & PolyProtect & 88.46\% & 5.90 & 0.40 & 7.10\\

& & MRL & 88.33\% & 2.42 & 1.22 & 5.29\\ 

& & MRL + PolyProtect & 88.29\% & 4.39 & 1.36 & 6.14\\ 

& & MRL + FHE & 88.23\% & 39.99 & 57.05 & 57.70\\ 

& & MRL + PolyProtect + FHE & 88.21\% & \textbf{40.47} & \textbf{57.12} & \textbf{57.72}\\ 

\hline
\end{tabular}
\end{center}
\end{table*}

\begin{table*}
\caption{Suppression Rate across different soft biometric attributes.}
\label{SuppressionRate}
\begin{center}
\begin{tabular}{|c|c|c|c|c|c|c|}
\hline
\textbf{Embeddings} & \textbf{Dataset} & \textbf{Template Protection} & \textbf{Identification Accuracy}&\textbf{Gender Attribute} & \textbf{Age Attribute} & \textbf{Ethnicity Attribute}\\
\hline
\multirow{6}{*}{FaceNet} & \multirow{6}{*}{CelebSet} &  PolyProtect & 99.42\%& 0.0073 & 0.0306 & 0.0075\\ 

& & MRL & 97.93\% & 0.0012 & 0.0206 & 0.0145\\ 

& & MRL + PolyProtect &97.00\% & 0.0183 & 0.0041 & 0.0037\\ 

& & MRL + FHE & 97.83\% & \textbf{0.4678} & \textbf{0.9302} & \textbf{0.9189}\\ 

& & MRL + PolyProtect + FHE & 96.95\% & \textbf{0.4678} & \textbf{0.9302} & 0.9187\\ 
\cline{2-7}

 & \multirow{6}{*}{BFW}  & PolyProtect & 85.04\% & 0.0019 & 0.0046 & 0.0038\\ 

& & MRL &84.42\%& 0.0339 & 0.0044 & 0.0035\\ 

& & MRL + PolyProtect &84.31\% & 0.0551 & 0.0689 & 0.0464\\ 

& & MRL + FHE &84.38\%& 0.4753 & 0.7034 & 0.7389\\ 

& & MRL + PolyProtect + FHE & 84.28\% & \textbf{0.4803} & \textbf{0.7053} & \textbf{0.7390}\\

\hline

\multirow{6}{*}{AdaFace} & \multirow{6}{*}{CelebSet} & PolyProtect &99.38\%& 0.0122 & 0.0008 & -0.0026\\

& & MRL & 97.95\% & -0.0072 & 0.0526 & -0.0273\\ 

& & MRL + PolyProtect & 97.59\% & 0.0011 & 0.0124 & -0.0201\\ 

& & MRL + FHE & 97.91\% & \textbf{0.4613} & 0.9322 & 0.9167\\ 

& & MRL + PolyProtect + FHE & 97.55\% & \textbf{0.4613} & \textbf{0.9323} & \textbf{0.9168}\\ 
\cline{2-7}

 & \multirow{6}{*}{BFW} & PolyProtect & 88.46\% & 0.0656 & 0.40 & 0.0869\\

& & MRL & 88.33\% & 0.0269 & 0.0143 & 0.0647\\ 

& & MRL + PolyProtect & 88.29\% & 0.0488 & 0.0159 & 0.0751\\ 

& & MRL + FHE & 88.23\% & 0.4444 & 0.6710 & 0.7061\\ 

& & MRL + PolyProtect + FHE & 88.21\% & \textbf{0.4498} & \textbf{0.6718} & \textbf{0.7063}\\ 

\hline
\end{tabular}
\end{center}
\end{table*}

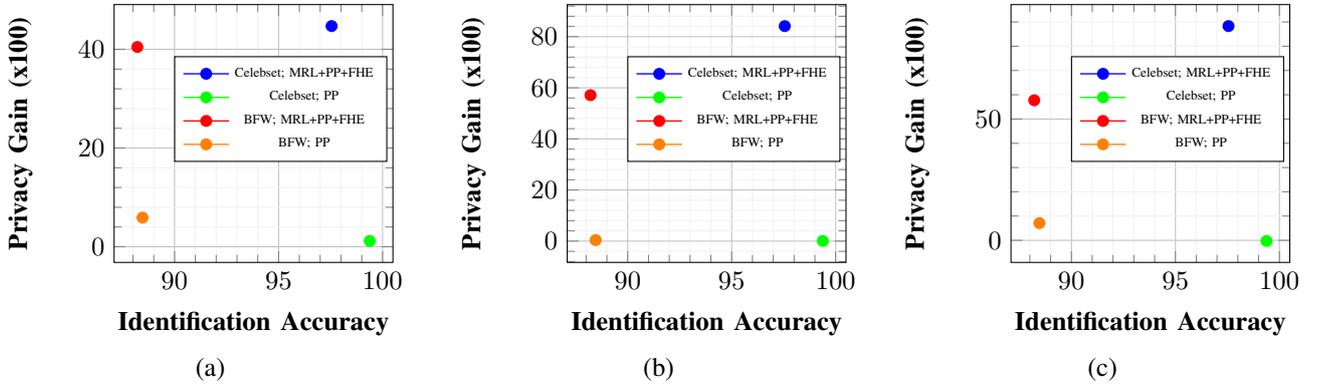
\begin{figure*}[!ht]
\begin{tabular}{ccc}

        \centering
    \begin{tikzpicture}
    \begin{axis}[
      width=5.25cm,
      height=5cm,
      xlabel={Identification Accuracy},
      ylabel={Privacy Gain (x100)},
      font=\bfseries,
      legend pos=north east,
      legend style={font=\tiny, at={(0.98,0.80)}, anchor=north east},
      grid=both,
      grid style={line width=.1pt, draw=gray!10},
      major grid style={line width=.2pt,draw=gray!50},
      minor tick num=4,
    ]

    % Scatter plots and lines
    \addplot[mark=*, blue, line width = 0.5pt] table[x=Acc, y=Gain] {
      Gain Acc
      44.71 97.55
    };
    \addlegendentry{Celebset; MRL+PP+FHE}

    \addplot[mark=*, green, line width = 0.5pt] table[x=Acc, y=Gain] {
      Gain Acc
      1.18 99.38
    };
    \addlegendentry{Celebset; PP}

    \addplot[mark=*, red, line width = 0.5pt] table[x=Acc, y=Gain] {
      Gain Acc
      40.47 88.21
    };
    \addlegendentry{BFW; MRL+PP+FHE}

    \addplot[mark=*, orange, line width = 0.5pt] table[x=Acc, y=Gain] {
      Gain Acc
      5.90 88.46
    };
    \addlegendentry{BFW; PP}
    
    \end{axis}
  \end{tikzpicture}
  \label{overlap_pp}
&
        \begin{tikzpicture}
    \begin{axis}[
      width=5.25cm,
      height=5cm,
      xlabel={Identification Accuracy},
      ylabel={Privacy Gain (x100)},
      font=\bfseries,
      legend pos=north east,
      legend style={font=\tiny, at={(0.98,0.80)}, anchor=north east},
      grid=both,
      grid style={line width=.1pt, draw=gray!10},
      major grid style={line width=.2pt,draw=gray!50},
      minor tick num=4,
    ]

    % Scatter plots and lines
    \addplot[mark=*, blue, line width = 0.5pt] table[x=Acc, y=Gain] {
      Gain Acc
      84.14 97.55
    };
    \addlegendentry{Celebset; MRL+PP+FHE}

    \addplot[mark=*, green, line width = 0.5pt] table[x=Acc, y=Gain] {
      Gain Acc
      0.07 99.38
    };
    \addlegendentry{Celebset; PP}

    \addplot[mark=*, red, line width = 0.5pt] table[x=Acc, y=Gain] {
      Gain Acc
      57.12 88.21
    };
    \addlegendentry{BFW; MRL+PP+FHE}

    \addplot[mark=*, orange, line width = 0.5pt] table[x=Acc, y=Gain] {
      Gain Acc
      0.40 88.46
    };
    \addlegendentry{BFW; PP}
    \end{axis}
  \end{tikzpicture}
  &
  \begin{tikzpicture}
    \begin{axis}[
      width=5.25cm,
      height=5cm,
      xlabel={Identification Accuracy},
      ylabel={Privacy Gain (x100)},
      font=\bfseries,
      legend pos=north east,
      legend style={font=\tiny, at={(0.98,0.80)}, anchor=north east},
      grid=both,
      grid style={line width=.1pt, draw=gray!10},
      major grid style={line width=.2pt,draw=gray!50},
      minor tick num=4,
    ]

    % Scatter plots and lines
    \addplot[mark=*, blue, line width = 0.5pt] table[x=Acc, y=Gain] {
      Gain Acc
      88.30 97.55
    };
    \addlegendentry{Celebset; MRL+PP+FHE}

    \addplot[mark=*, green, line width = 0.5pt] table[x=Acc, y=Gain] {
      Gain Acc
      -0.25 99.38
    };
    \addlegendentry{Celebset; PP}

    \addplot[mark=*, red, line width = 0.5pt] table[x=Acc, y=Gain] {
      Gain Acc
      57.72 88.21
    };
    \addlegendentry{BFW; MRL+PP+FHE}

    \addplot[mark=*, orange, line width = 0.5pt] table[x=Acc, y=Gain] {
      Gain Acc
      7.10 88.46
    };
    \addlegendentry{BFW; PP}
    \end{axis}
  \end{tikzpicture}\\
    (a) & (b) & (c)\\
    
        \end{tabular}
        \caption{Privacy Gain of our proposed approach compared to baseline (PP) across different attributes - (a) Gender, (b) Age and (c) Ethnicity - using \textbf{Adaface} (MRL - Matryoksha Representation Learning; FHE - Fully Homomorphic Encryption; PP - PolyProtect).}
        \label{Fig:adaFaceResults}
     
\end{figure*}

\begin{figure*}
\begin{tabular}{ccc}

        \centering
    \begin{tikzpicture}
    \begin{axis}[
      width=5.25cm,
      height=5cm,
      xlabel={Identification Accuracy},
      ylabel={Privacy Gain (x100)},
      font=\bfseries,
      legend pos=north east,
      legend style={font=\tiny, at={(0.98,0.80)}, anchor=north east},
      grid=both,
      grid style={line width=.1pt, draw=gray!10},
      major grid style={line width=.2pt,draw=gray!50},
      minor tick num=4,
    ]

    % Scatter plots and lines
    \addplot[mark=*, blue, line width = 0.5pt] table[x=Acc, y=Gain] {
      Gain Acc
      45.90 96.95
    };
    \addlegendentry{Celebset; MRL+PP+FHE}

    \addplot[mark=*, green, line width = 0.5pt] table[x=Acc, y=Gain] {
      Gain Acc
      0.72 99.42
    };
    \addlegendentry{Celebset; PP}

    \addplot[mark=*, red, line width = 0.5pt] table[x=Acc, y=Gain] {
      Gain Acc
      45.75 84.28
    };
    \addlegendentry{BFW; MRL+PP+FHE}

    \addplot[mark=*, orange, line width = 0.5pt] table[x=Acc, y=Gain] {
      Gain Acc
      0.18 85.04
    };
    \addlegendentry{BFW; PP}
    
    \end{axis}
  \end{tikzpicture}
  \label{overlap_pp}
&
        \begin{tikzpicture}
    \begin{axis}[
      width=5.25cm,
      height=5cm,
      xlabel={Identification Accuracy},
      ylabel={Privacy Gain (x100)},
      font=\bfseries,
      legend pos=north east,
      legend style={font=\tiny, at={(0.98,0.80)}, anchor=north east},
      grid=both,
      grid style={line width=.1pt, draw=gray!10},
      major grid style={line width=.2pt,draw=gray!50},
      minor tick num=4,
    ]

    % Scatter plots and lines
    \addplot[mark=*, blue, line width = 0.5pt] table[x=Acc, y=Gain] {
      Gain Acc
      81.56 96.95
    };
    \addlegendentry{Celebset; MRL+PP+FHE}

    \addplot[mark=*, green, line width = 0.5pt] table[x=Acc, y=Gain] {
      Gain Acc
      2.68 99.42
    };
    \addlegendentry{Celebset; PP}

    \addplot[mark=*, red, line width = 0.5pt] table[x=Acc, y=Gain] {
      Gain Acc
      66.58 84.28
    };
    \addlegendentry{BFW; MRL+PP+FHE}

    \addplot[mark=*, orange, line width = 0.5pt] table[x=Acc, y=Gain] {
      Gain Acc
      0.43 85.04
    };
    \addlegendentry{BFW; PP}
    \end{axis}
  \end{tikzpicture}
  &
  \begin{tikzpicture}
    \begin{axis}[
      width=5.25cm,
      height=5cm,
      xlabel={Identification Accuracy},
      ylabel={Privacy Gain (x100)},
      font=\bfseries,
      legend pos=north east,
      legend style={font=\tiny, at={(0.98,0.80)}, anchor=north east},
      grid=both,
      grid style={line width=.1pt, draw=gray!10},
      major grid style={line width=.2pt,draw=gray!50},
      minor tick num=4,
    ]

    % Scatter plots and lines
    \addplot[mark=*, blue, line width = 0.5pt] table[x=Acc, y=Gain] {
      Gain Acc
      90.78 96.95
    };
    \addlegendentry{Celebset; MRL+PP+FHE}

    \addplot[mark=*, green, line width = 0.5pt] table[x=Acc, y=Gain] {
      Gain Acc
      0.75 99.42
    };
    \addlegendentry{Celebset; PP}

    \addplot[mark=*, red, line width = 0.5pt] table[x=Acc, y=Gain] {
      Gain Acc
      67.97 84.28
    };
    \addlegendentry{BFW; MRL+PP+FHE}

    \addplot[mark=*, orange, line width = 0.5pt] table[x=Acc, y=Gain] {
      Gain Acc
      0.35 85.04
    };
    \addlegendentry{BFW; PP}
    \end{axis}
  \end{tikzpicture}\\
    (a) & (b) & (c)\\
    
        \end{tabular}
        \caption{Privacy Gain of our proposed approach compared to baseline (PP) across different attributes - (a) Gender, (b) Age and (c) Ethnicity - using \textbf{FaceNet} (MRL - Matryoksha Representation Learning; FHE - Fully Homomorphic Encryption; PP - PolyProtect).}
        \label{Fig:faceNetResults}
     
\end{figure*}
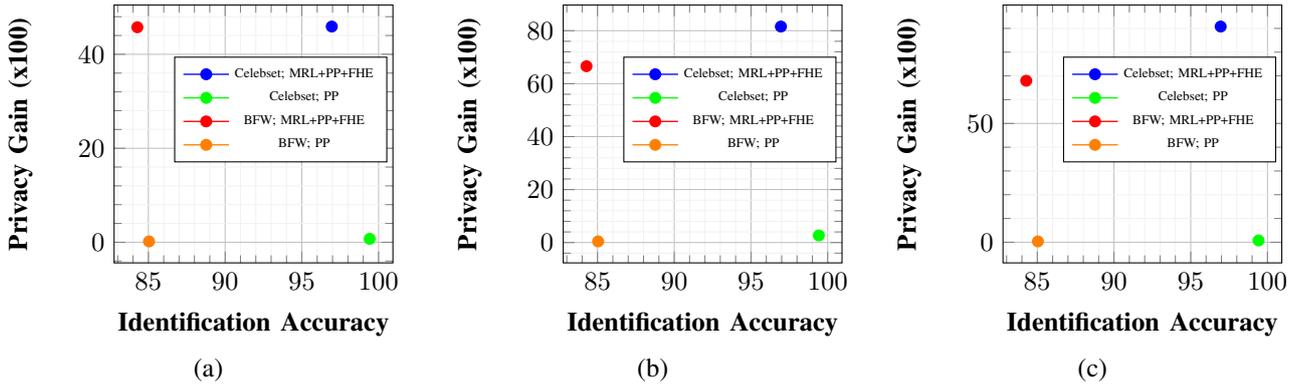

\section{Results}
As evident from Tables \ref{Accuracy}, \ref{PrivacyGain}, \ref{SuppressionRate} our proposed approach prevents the leakage of soft biometrics from face embeddings with minimal loss in identification accuracy  ($<$2.5\%), high Privacy Gain and high Suppression Rate. We can observe that our approach reduced the classification accuracies of soft biometric attributes to the level of random chance across the two datasets. We could also achieve an almost ideal Privacy Gain and Suppression Rate in certain scenarios, but we believe this could be because of the imbalanced nature of our datasets. Our experiments showcase that a combination of MRL, FHE, and PolyProtect in this order yields maximum protection against soft biometric leakage (Fig. \ref{Fig:adaFaceResults} and Fig. \ref{Fig:faceNetResults}).

The consistent efficacy of our method across two different embeddings (FaceNet and AdaFace) and datasets (CelebSet and BFW) shows the ability of our method to generalize in different conditions. We also prove the ability of FHE to work seamlessly with template protection schemes and embedding compression techniques for preserving the privacy of face templates. 

Additionally, we find that utilizing a soft biometrics classifier network trained on plaintext data allows for seamless inference within the FHE domain, yielding the same predictive performance as that in the unencrypted domain (as depicted in Table \ref{Accuracy} in "None"). This underscores the feasibility of conducting not only identification but also soft biometric analysis in the secure FHE domain.

%Additional experiments, as in Figure \ref{Fig:ablation}, with the different paramters of PolyProtect show us that it impacts Gender and Age prediction performance. We observe that increasing overlap resulted in higher accuracy for both Gender and Age prediction. This is in agreement with the results from \cite{base} which suggest that increasing the overlap of PolyProtecthelps the transformation retain more information thereby improving the identification accuracy. While this is good in terms of accuracy, it is not good for privacy. Hence, the use of FHE alleviates the issues with privacy while retaining higher accuracy. 

\section{Conclusion}
In this paper, we propose to use FHE in combination with template protection and compression to secure the face template and prevent soft biometric leakage. We show that soft biometric attributes from face embeddings can be strictly protected while preserving identification accuracy. In our approach, we compress the face embeddings using MRL (Matryoksha Representation Learning), encrypt them, and then apply PolyProtect as the template protection scheme. The identification performance of the encrypted template compared with the unencrypted version is unchanged. Since FHE guarantees are based on strong theoretical principles, privacy and security are ensured, and only authorized individuals with the secret key will be able to access the results from the FHE computation. 

\noindent\textit{This material is based upon work supported by the Center for Identification Technology Research and the National Science Foundation under Grant Nos. 1841517 and 1822190.}

%%%%%%%%%%%%%%%%%%%%%%%%%%%%%%%%%%%%%%%%%%%%%%%%%%%%%%%%%%%%%%%%%%%%%%%%%%%%%%%%

%%%%%%%%%%%%%%%%%%%%%%%%%%%%%%%%%%%%%%%%%%%%%%%%%%%%%%%%%%%%%%%%%%%%%%%%%%%%%%%%

%%%%%%%%%%%%%%%%%%%%%%%%%%%%%%%%%%%%%%%%%%%%%%%%%%%%%%%%%%%%%%%%%%%%%%%%%%%%%%%%

%%%%%%%%%%%%%%%%%%%%%%%%%%%%%%%%%%%%%%%%%%%%%%%%%%%%%%%%%%%%%%%%%%%%%%%%%%%%%%%%

%\newpage
{
\bibliographystyle{unsrt}
\bibliography{FG2024}
}

\end{document}